\documentclass[amsmath,amssymb,aps,prl,nofootinbib,twocolumn]{revtex4-2}
\usepackage{graphicx}
\usepackage{dcolumn}
\usepackage{bm}
\usepackage{hyperref}
\usepackage{color}
\usepackage{bbm}

\hypersetup{
colorlinks=true,         
linkcolor=blue,          
citecolor=red,        
urlcolor=red            
}

\newcommand{\co}{\nu}

\begin{document}

\title{Through a Black Hole Singularity}

\author{Flavio Mercati$^{1}$ and David Sloan$^2$}

\affiliation{
\vspace{11pt}
$^1$Departamento de F\'isica, Universidad de Burgos, E-09001 Burgos, Spain;
\\
$^2$Department of Physics, Lancaster University, Lancaster UK.}

\date{\today}

\begin{abstract}
\noindent 
We show that the Kantowski--Sachs model of a Schwarzschild black hole interior can be slightly generalized in order to accommodate spatial metrics of different orientations, and in this formulation the equations of motion admit a variable redefinition that makes the system regular at the singularity. This system will then traverse the singularity in a deterministic way (information will be conserved through it), and evolve into a time-reversed and orientation-flipped Schwarzschild white hole interior.
\end{abstract}

\maketitle

The interior of the Schwarzschild metric can be understood as a special case of the Kantowski--Sachs class of spacetimes~\cite{Kantowski1966}. These are homogeneous cosmological models with an $S^2 \times \mathbb{R}$ spatial topology, and a spacetime metric of the form:
\begin{equation}\label{KSansatz}
d\tau^2 = - N(\sigma)^2 d\sigma^2 + A(\sigma)^2 d \rho^2 + B(\sigma)^2 d\Omega^2 \,.
\end{equation}
The ordinary Schwarzschild metric is found as the particular case
\begin{equation} \label{Schw}
\textstyle
N= \left(\frac{2M}{\sigma} -1\right)^{-1/2} \,, ~~ 
A= \left(\frac{2M}{\sigma} -1\right)^{1/2} \,, ~~ 
B = \sigma \,,
\end{equation}
if we call $\sigma = r$ and $\rho = t$. This, of course, is only valid when $r < 2M$, \emph{i.e.} the region inside the event horizon, where the $r$ coordinate is timelike and the $t$ coordinate is spacelike.

The manipulation we described allows us to understand the Schwarzschild singularity in a similar manner to the Big Bang, and to translate progress in the understanding of homogeneous cosmological singularities into advancement in the physics of black holes. A recent approach~\cite{ThroughTheBigBang,DavesScalarpaper,FlaviosInflationpaper}\footnote{See also~\cite{Kamenshchik2017} for a closely-related approach, and~\cite{Bonanno2017} for a possible quantum origin for the quiescence mechanism.}  allowed us to prove an existence and uniqueness theorem for the solutions of Einstein's equations at and beyond the singularity of homogeneous cosmologies. By extending the configuration space of the theory to include the information about the orientation of spatial slices, this result allows us to prove that to each and every collapsing solution ending up in a singularity, there corresponds one and only one expanding solution that evolves away from the singularity with opposite orientation. The singularity is a degenerate hypersurface which cannot support a nonzero volume because it is effectively one- or two-dimensional. At this hypersurface, the spatial orientation flips.

Thanks to the device described at the beginning, these insights on homogeneous cosmologies can be applied to black hole singularities described as Kantowski--Sachs spacetimes.

\section{The empty Kantowski--Sachs model}

After imposing the ansatz~(\ref{KSansatz}), the Einstein--Hilbert Lagrangian reads
\begin{equation}\label{EinsteinHilbertLagrangian}
\begin{aligned}
 L =&  {\textstyle \frac{1}{2 \kappa}} \int d^3 x \sqrt{-g} R =
\textstyle \frac{4 \pi N \lambda}{\kappa} \left(  A - \frac{A (\dot B)^2 +  2 B \dot A \dot B}{N^2}\right) + \dot K  \,,
\end{aligned}
\end{equation}
where $\kappa = 8 \pi G c^{-4}$ and $\lambda = \int_{r_1}^{r_2} d \rho $ is the width of a fiducial interval of radii over which we integrate (by homogeneity, the metric outside this interval will be identical to the one inside). $K = \frac{4 \pi \lambda}{\kappa N} \left( \dot A B^2 + 2 \dot B A B \right)$ appears as a total derivative, and is therefore a boundary term that can be removed (it is minus the Gibbons--Hawking--York term~\cite{York72,Gibbons-Hawking}).

 In terms of the canonical momenta $P_A = \partial L / \partial \dot A$, $P_B = \partial L / \partial \dot B$, we can write the total Hamiltonian $H = P_A \dot A +P_B \dot B - L$ as:
\begin{equation}
H = \frac{N}{\co^2} \left(
\frac{P_A^2 A - 2 P_A P_B B}{4 B^2}
-\co^4 A \right) \,,
\end{equation}
where $\co =  \sqrt{\frac{4 \pi \lambda}{\kappa}}$. With the following canonical transformation:
\begin{equation}\label{ABtoxy}
\begin{gathered}
A = \frac{e^{ - \frac x {\sqrt 2}}}{\co} \,, ~~
B = \frac{e^{  \frac {x+y} {\sqrt 2}}}{\co} \,, 
\\
P_A = -\sqrt{2} \co e^{\frac{x}{\sqrt{2}}} (p_x - p_y) \,, ~~
P_B = \sqrt{2} \co  e^{-\frac{x+y}{\sqrt{2}}} p_y \,,
\end{gathered}
\end{equation}
the Hamiltonian takes the simple form:
\begin{equation}\label{HamiltonianFreeKS}
H = \frac{N \, \co}{2}  e^{-\frac{x+2y}{\sqrt{2}}} \left(
p_x^2 - p_y^2 - 2 e^{\sqrt 2 y}  \right) \,.
\end{equation}

We are free to choose the lapse function $N$, and the obvious choice is $N = \frac 1 {\nu}  e^{\frac{x+2y}{\sqrt{2}}}$, which simplifies the prefactor and gives us the elementary Hamiltonian
\begin{equation}
H = \frac{1}{2}   \left(
p_x^2 - p_y^2 \right) -  e^{\sqrt 2 y}  \,.
\end{equation}
This Hamiltonian makes $p_x$ a conserved quantity, and the Hamiltonian constraint $H \approx 0$ imposes that  $ p_x^2 = p_y^2  + 2  e^{\sqrt 2 y}$. This is the Hamiltonian of a one-dimension nonrelativistic point particle with potential $2  e^{\sqrt 2 y}$ and energy $p_x^2$. The general solution to Hamilton's equations is:
\begin{equation}\label{SolutionEmptyK-S}
\begin{gathered}
x = p_x (s- s_1)   \,, ~~ p_y = k \tanh \left( \frac{k(s-s_2)}{\sqrt 2} \right) \,,\\
y = - \textstyle  {\sqrt 2} \log  \left[ \frac{\sqrt{2}}{|k|}   \cosh \left(  \frac{k (s - s_2 )}{\sqrt{2}}  \right) \right]   \,,
\end{gathered}
\end{equation}
where $p_x$ now is a constant of motion, and $k$, $s_1$, and $s_2$ are constants of integration. Moreover, the Hamiltonian constraint imposes that $p_x^2 = k^2$.  The asymptotic components of the velocity are $\dot x = p_x$ and $\dot y \xrightarrow[s \to \pm \infty]{} \mp  | k | =  \mp  |p_x|$. In the $x-y$ plane, all solutions look like a ball bouncing off an exponential slope and rolling inertially to infinity at a 45$^\circ$ angle.
\begin{figure}[t!]
\center
\includegraphics[width=0.47\textwidth,trim=0 2 1 2,clip]{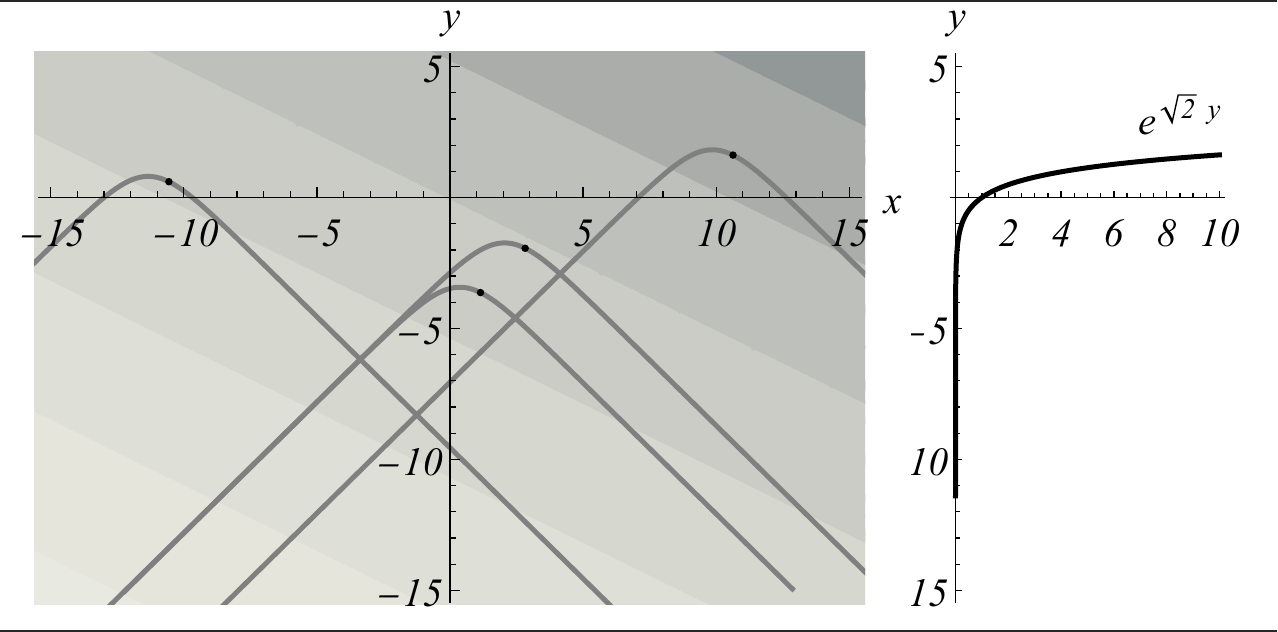}
\caption{Examples of solutions of the Kantowski--Sachs system in the $x-y$ plane, with a plot of the potential on the side. The curves asymptotes to the singularity on the left, and the horizon on the right. The gradient on the background represents the values of the volume degree of freedom $v$ (darker = larger), and the black dots represent the points of maximum volume.}\label{EmptySolutions}
\end{figure}

From Fig.~\ref{EmptySolutions} we can see that the volume $v \propto A B^2$ of our fiducial region is convex (as a function of $s$), going to zero as $s \to \pm \infty$, and reaching a unique maximum in between.

We can calculate the Ricci tensor on the solution~(\ref{SolutionEmptyK-S}), and only one  component turns out to be nonzero:
\begin{equation}\label{RicciTensor}
R_{\mu\nu} = \left( k^2 - p_x^2 \right) \delta^0{}_\mu \delta^0{}_\nu  \,,
\end{equation}
and if we impose the Hamiltonian constraint $p_x^2 = k^2$, the spacetime we get is Ricci-flat.  To highlight the location of the singularity, we can calculate the Kretschmann scalar:
\begin{equation}
R_{\mu\nu\rho\sigma} R^{\mu\nu\rho\sigma} = 
\frac{3 \co^4 \left( e^{\sqrt{2} k (s_2 - s)}+1\right)^6}{k^4 e^{ - \sqrt{2}  k (2 s_1 + 4 s_2 )} } \,, 
\end{equation}
and see that it diverges when $\text{sign}(k) s \to - \infty$.

\begin{figure}[t!]
\center
\includegraphics[width=0.5\textwidth]{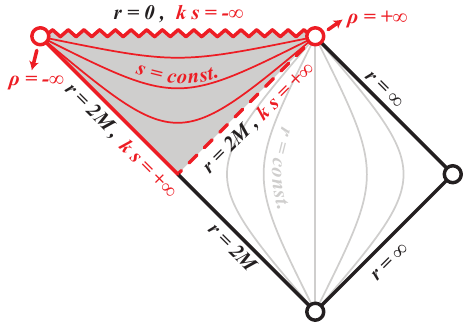}
\caption{The shaded region corresponds to the patch of Schwarzschild spacetime that is covered by the $s$, $\rho$ coordinates as it appears in the Penrose--Carter diagram. The borders of this coordinate patch are represented by the two red dots, the singularity and the horizon.}\label{PenroseDiagramFig1}
\end{figure}

\section{Obtaining the Schwarzschild metric}

The Schwarzschild solution can be obtained  by setting\footnote{In this Section we assume that $p_x = k$, the other case $p_x = -k$ can be straightforwardly worked out analogously.}
\begin{equation} \label{ConstraintOnIntegrationConstants} \textstyle
k (s_1-s_2)  = \sqrt{2} \log \co \,,
\end{equation}
then one can see that $A$ and $B$, expressed in terms of the solution $x$, $y$ of~(\ref{SolutionEmptyK-S}) through the relations~(\ref{ABtoxy}), satisfy the equation
\begin{equation}\textstyle
A^2 =  \frac{2M}{B}  - 1 \,,
\end{equation}
where, as it turns out, $2 M  =\frac{\sqrt{2} |k|}{\co^2}$. We can recover the full Schwarzschild metric by making a time reparametrization $s \to r$ that transforms $B[s(r)] = r$ (which is legitimate because, as is easy to check, on-shell $\dot B$ is definite, and $B$ is therefore monotonic), which gives 
\begin{equation} \label{TimeReparametrization} \textstyle
s = s_2 - \frac{1}{\sqrt 2 k} \log \left( \frac{2M}{r} -1 \right) \,,
\end{equation}
and transforms the lapse into 
\begin{equation}\textstyle
N[s(r)] \frac{\partial s}{\partial r} = \left( \frac{\sqrt{2}|k|}{\co^2 r}  - 1 \right)^{-1/2} = \left( \frac{2M}{r}  - 1 \right)^{-1/2}  \,.
\end{equation}
Notice that \emph{all} solutions~(\ref{SolutionEmptyK-S}) represent a Schwarzschild spacetime. Those whose integration constants fail to satisfy~(\ref{ConstraintOnIntegrationConstants}) are just associated to a rescaled metric:
\begin{equation}\textstyle
N = \alpha \left( \frac{2M}{r} -1 \right)^{-1/2}
\,, ~~
A = \alpha \left( \frac{2M}{r} -1 \right)^{1/2}
\,, ~~
B= \alpha r \,,
\end{equation}
where $\alpha = \nu\,e^{-\frac{k (s_1 - s_2) - x_0}{\sqrt 2} } $, and therefore a redefinition of units can reabsorb this.

The time redefinition~(\ref{TimeReparametrization}) gives us another way to identify the values of $s$ corresponding to the singularity (which is at $r \to 0$).  The reparametrization monotonically maps $r \in (0,2M)$ into $\text{sign}(k) s \in (-\infty , \infty)$. The singularity $r \to 0^+$ coincides with $\text{sign}(k) s \to - \infty$, \emph{i.e.} when $x \to - \infty$. The other limit  $x \to + \infty$ coincides with the horizon $r \to (2M)^-$.

\section{Shape space and orientation}

One linear combination of the $x$ and $y$ variables corresponds to the scale degree of freedom, while the other is conformally invariant and determines the \emph{shape} of our spatial hypersurface (in particular, it determines the ratio between the radial extension of our coordinate patch and its areal radius). To disentangle scale and shape, consider the determinant of the spatial metric,  $\det g = A^2 B^4 = \nu^{-6} e^{\sqrt{2}(x+2y)}$, which is a pure scale degree of freedom. Therefore $x+2y$ determines the scale, while the orthogonal direction in the $(x,y)$ plane determines the shape. The following linear canonical transformation separates between scale $z$ and shape $w$:
\begin{equation}\label{CanonicalTransformToShapeVariables}
\begin{aligned}
&x =  \frac{1}{\sqrt{3}} \left( 2 w - z \right)\,,~~
y = \frac{1}{\sqrt{3}} \left(2 z - w \right) \,,~~
\\
&p_x =   \frac{p_z + 2 p_w}{\sqrt{3}} \,,~~
p_y = \frac{2 p_z + p_w}{\sqrt{3}}  \,,
\end{aligned}
\end{equation}
so that now $\det g = \nu^{-6} e^{\sqrt{6} z}$  depends on $z$ alone. In the new variables, the Hamiltonian constraint takes a simple form:
\begin{equation}\label{HamiltonianKS_ShapeVariables}
H = \textstyle \frac 1 2 \left( p_w^2  - 
p_z^2   \right)  - e^{\sqrt{\frac 2 3} (2 z -  w )}  \,,
\end{equation}
notice that, as usual in a constant-mean-extrinsic-curvature foliation, the scale degree of freedom gives a negative contribution to the kinetic term~\cite{FlavioSDbook}.

Notice now that the coordinate change from the $(w,z)$ variables to the original $(A,B)$ variables,
\begin{equation}
A = \frac{e^{ \frac{1}{\sqrt{6}} \left( z - 2 w \right)}}{\co} \,, ~~
B = \frac{e^{  \frac{1}{\sqrt{6}} \left( w + z \right) }}{\co} \,,
\end{equation}
is not surjective: it only maps $\mathbbm{R}^2$ to the first quadrant ($A>0,B>0$) of $\mathbbm{R}^2$. Normally this would not be a problem, because the metric~(\ref{KSansatz}) depends only on the square of $A$ and $B$, and the configuration space of Kantowski--Sachs metrics is more appropriately defined as the quotient of the $(A,B)$ plane by reflections of $A$ an $B$. However, there is a bit of information that is erased by this quotienting procedure, which we might want to keep track of instead. This is the \emph{orientation} of our spatial manifold, which is encoded, for example, in the triad formulation of the metric~\cite{EguchiBook}
\begin{equation}
g_{ij} = \delta_{ab} e^a{}_i e^b{}_j \,,
\end{equation}
the associated volume form $e^1 \wedge e^2 \wedge e^3 $ defines an orientation on our manifold. In this formulation, under the Kantowski--Sachs ansatz the frame field components are linear in $A$ and $B$, and the volume form reads $ e^1 \wedge e^2 \wedge e^3  = A \, B^2$. Therefore, the sign of $A$ determines the orientation of our spatial hypersurface. 

The variable $z$ parametrizes the scale degree of freedom, while $w$ determines the shape of our spatial manifold, and it makes sense to include the information regarding the orientation into the ``shape space'' of our model~\cite{FlavioSDbook,ThroughTheBigBang}. We can then extend the shape space, by defining two coordinate patches, $w_+ \in \mathbbm{R}$ and $w_- \in \mathbbm{R}$, which are mapped to the two possible signs of $A$:
\begin{equation}\label{CoordinatesWithOrientation}
A = 
\left\{\begin{aligned}
&~\frac{e^{ \frac{1}{\sqrt{6}} \left( z - 2 w_+ \right)}}{\co} \,, ~~ &\text{if } A>0  \,,
\\
&- \frac{e^{ \frac{1}{\sqrt{6}} \left( z - 2 w_- \right)}}{\co} \,, ~~ &\text{if } A<0 \,.
\end{aligned}
\right.
\end{equation}
The above map sends two copies of $\mathbbm{R}$ onto the two halves of the real line.

The two possible signs of $A$ correspond to the choice between left- and right-handed triads compatible with the metric, $e_\pm$. Taking the Schwarzschild solution as our guide we expect that as we approach the singularity $|A| \rightarrow \infty$. The singularity is potentially a point of transition between $e_+$ and $e_-$, hence a point at which the orientation of our space may change. By extending our description to the coordinate patches $w_\pm$ we allow for our dynamics to distinguish between orientations.

\begin{figure}[t!]
\center
\includegraphics[width=0.5\textwidth]{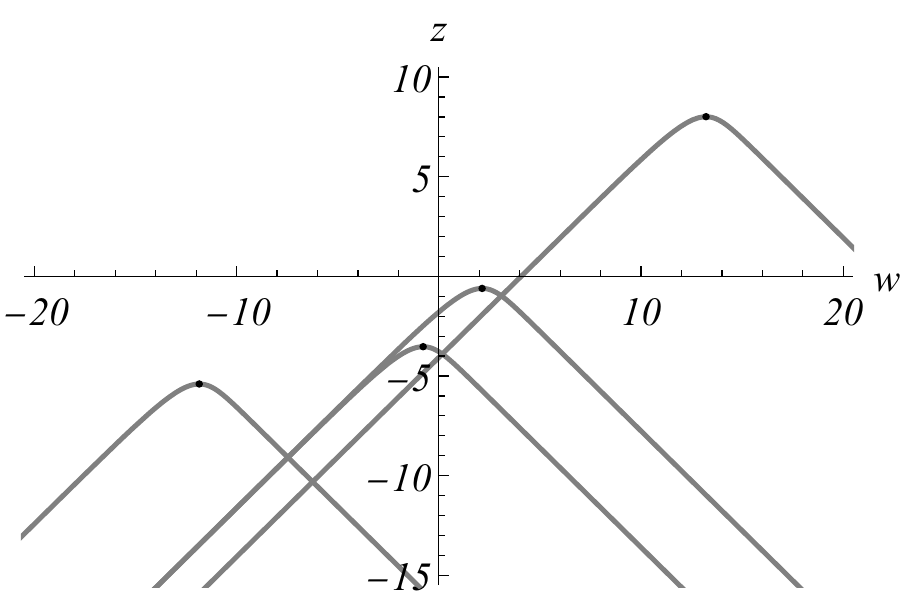}
\caption{The same solutions of Fig.~\ref{EmptySolutions}, this time plotted in the $w-z$ plane. The singularity is reached asymptotically as $w \to - \infty$, while the horizon is at $w \to + \infty$.}
\end{figure}

At this point we could propose a continuation theorem along the lines of what was done in Bianchi IX~\cite{ThroughTheBigBang,FlaviosInflationpaper,DavesScalarpaper}, however such a theorem would be, in the present case, trivial. This is because the theorem of~\cite{ThroughTheBigBang,FlaviosInflationpaper,DavesScalarpaper} depends on the presence of more than one shape degree of freedom, and it becomes trivial in the case of a one-dimensional shape space. In fact, at the core of the continuation result, is the fact that one can decouple the scale degree of freedom (which is singular at the singularity) from the shape ones, and express the dynamics as a differential system in which the change of one shape degree of freedom is expressed in terms of the change in the others. This is the fundamental idea behind the ``Shape
Dynamics'' formulation of General Relativity~\cite{FlavioSDbook}, and the papers~\cite{ThroughTheBigBang,FlaviosInflationpaper,DavesScalarpaper} show how this intrinsic dynamics of pure shapes is regular at the singularity and can be continued deterministically through it. However, when we have only one shape degree of freedom, its change cannot be expressed in terms of other shape degrees of freedom. The intrinsic shape dynamics reduces to the prediction of an unparametrized curve on a one-dimensional manifold (a circle), and there is only one such curve. The fact that this curve continues through the singularity (which is located at a particular point on the circle) is a trivial statement.

For this reason, we are compelled to add some more shape degrees of freedom, in order to have a shape space of dimension at least two, where the fact that the intrinsic shape dynamics continues uniquely through the singularity is a nontrivial statement. The simplest way to do this is to add a homogeneous scalar field, which contributes with one shape degree of freedom. Notice that in~\cite{ThroughTheBigBang,FlaviosInflationpaper}
too we were forced to add (at least) one scalar field, but for a different reason. In fact, in these papers we were interested in the Bianchi IX cosmological model, which already comes equipped with a two-dimensional shape space. However, unless a stiff matter source is added, this model has an essential singularity at the big bang,  which makes continuation impossible. The the simplest form of stiff matter is a scalar field without mass or potential, the addition of which causes the system to transition to a state that is known as ``quiescence'', after which the dynamics ceases to be chaotic and admits a deterministic continuation through the singularity. In the present case, we add the scalar field just because we need additional scale degrees of freedom and that is the simplest option. The dynamics of the Kantowski--Sachs model can be continued through the singularity independently of the presence of scalar fields or stiff matter sources, because it is not chaotic like Bianchi IX.

\section{Homogeneous scalar field}\label{ScalarFieldSec}

To include a homogeneous scalar field to the Einstein--Hilbert Lagrangian~(\ref{EinsteinHilbertLagrangian}) we need to add the following term:
\begin{equation}
\begin{aligned}
L_\varphi &= -\int d^3 x \sqrt{-g} \left[ \frac 1 2 g^{\mu\nu} \partial_\mu \varphi \partial_\nu \varphi + V(\varphi) \right] = 
\\
&=   4 \pi \lambda \,  A \, B^2 \left[\frac 1 2  N^{-1} (\dot \varphi)^2 - N \, V(\varphi) \right] \,.
\end{aligned}
\end{equation}
Notice that the homogeneous ansatz for the scalar field corresponds, in the limit $k s \to \infty$ ($r \to (2M)^-$), to a field that is constant on the horizon. This could be taken as the s-wave contribution in an expansion in spherical harmonics around a Schwarzschild background.
We can now show how the Hamiltonian~(\ref{HamiltonianFreeKS}) generalizes in presence of a minimally-coupled homogeneous scalar field $\varphi$ (with the convenient choice of lapse $N = \frac 1 {\nu}  e^{\frac{x+2y}{\sqrt{2}}}$):
\begin{equation}\label{HamiltonianKS+Scalar}
H = \textstyle \frac 1 2  \left(
p_x^2 - p_y^2 + \frac 1 \kappa \pi_\varphi^2 \right) +U(x,y,\varphi) \,,
\end{equation}
where $\pi_\varphi$ is the momentum canonically conjugate to $\varphi$, and $U(x,y,\varphi)= - e^{\sqrt 2 y}  + \frac{\kappa}{\co^2 } e^{\sqrt{2}(x+2y)} V(\varphi)$ is the sum of the geometric and the scalar field potentials.

The first thing to notice is that the potential term breaks the conservation of the momentum $p_x$, and is capable of making the variable $x$ non-monotonic, potentially preventing it from reaching the singularity $x \to - \infty$. However, under certain not particularly restrictive conditions on the form of $V(\varphi)$,\footnote{Essentially, $V(\varphi)$ can go to infinity as $\varphi \to \pm \infty$, but it has to do so slower than $\exp(|\varphi|^{1+\epsilon})$ at least in one direction~\cite{FlaviosInflationpaper}.} one can see that there will be large classes of solutions in which $e^{\sqrt{2}(x+2y)}V(\varphi)$ asymptotes to zero, and $x$ and $y$ asymptote to the straight-line motion that ends in the singularity at $x \to - \infty$. This argument traces closely the more in-depth discussion developed in~\cite{FlaviosInflationpaper} with regards to scalar field (and inflationary) potentials.

We are interested in solutions that reach the singularity, and, by what we have just observed, these are such that the scalar field asymptotes to free dynamics (the potential $V(\varphi)$ becomes negligible near the singularity), and the solutions are identical to Eqs.~(\ref{EinsteinHilbertLagrangian}), with the addition of $\varphi = p_u \, (s - s_3)$, $p_u = \text{\it const.}$. What changes is the form of the Hamiltonian constraint:
\begin{equation}\label{HamConstWithPhi}
p_x^2 + \frac 1 \kappa \pi_\varphi^2 -k^2  = 0 \,, 
\end{equation}
this implies that the asymptotic motion in the $x-y$ plane is not at $45^\circ$, but at a steeper angle. Eq.~(\ref{HamConstWithPhi}) has another consequence: the Ricci tensor vanishes only when $\pi_\varphi=0$. It is only in absence of the scalar field that spacetime is Ricci-flat, and isometric to the Schwarzschild metric.

\section{Shape space with orientation,\\
and its compactification}

\begin{figure}[t!]
\center
\includegraphics[width=0.5\textwidth]{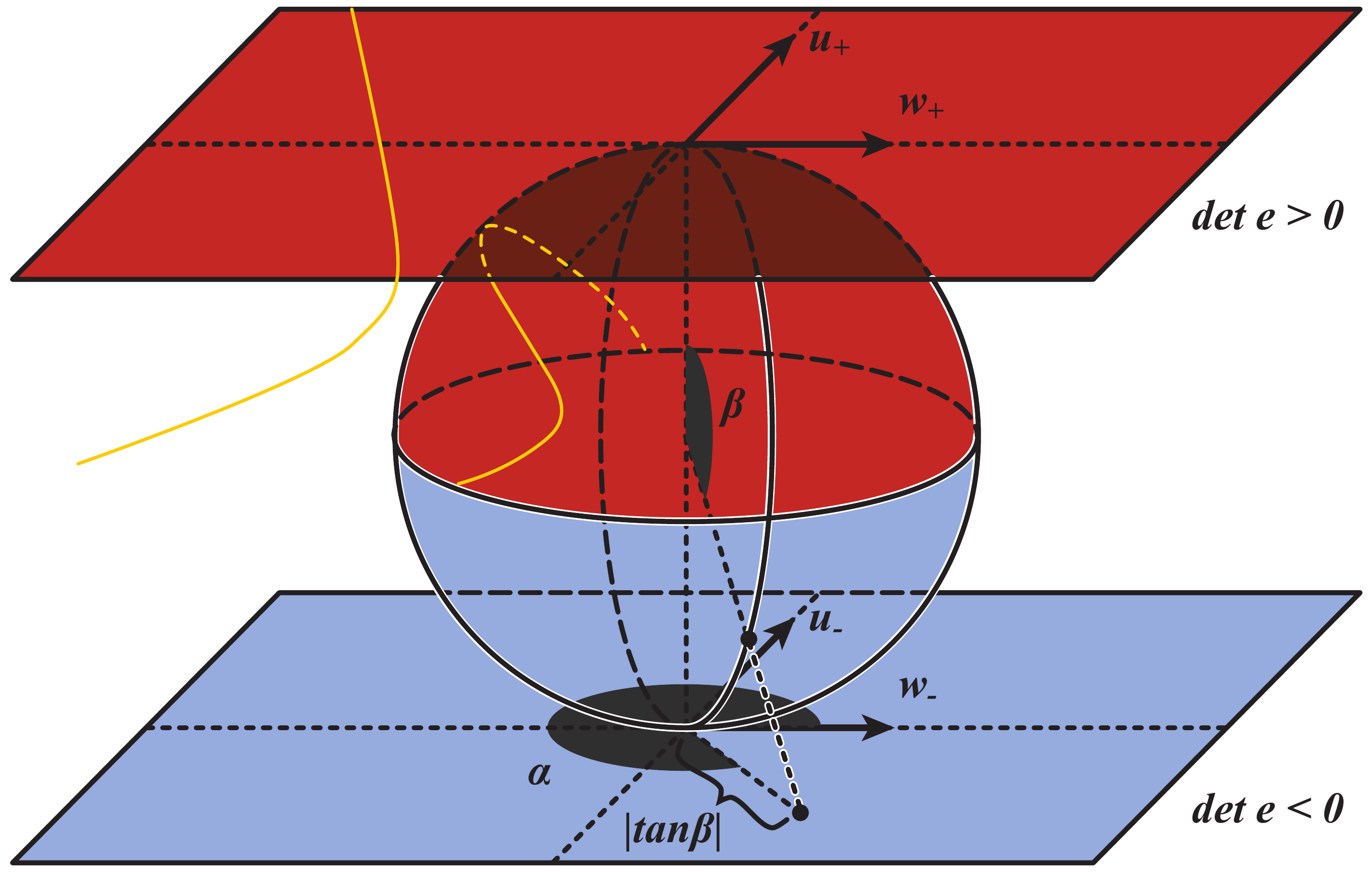}\label{GnomonicFig}
\caption{Shape space with orientation: each hemisphere represents an orientation, and each point on the sphere represents different values of the shape degrees of freedom $(w,u)$. The poles coincide with the value $u = w =0$, while the equator corresponds to the border of the $(w,u)$ plane at infinity. A solution curve is shown on the top plane, together with its projection on the northern hemisphere.}
\end{figure}

First, it is convenient to change the scalar field variable $\varphi$ to a dimensionless one, by means of the following canonical transformation:
\begin{equation}
u = \sqrt{\kappa} \, \varphi \,, ~~~ p_u = \pi_\varphi / \sqrt{\kappa} \,.
\end{equation}
Then, we can repeat the transformation~(\ref{CanonicalTransformToShapeVariables}) in order to separate scale and shape degrees of freedom. In the new variables, the Hamiltonian constraint takes this form:
\begin{equation}\label{HamiltonianKS+Scalar2}
H = \textstyle \frac 1 2 \left( p_w^2 +  p_u^2  - 
p_z^2   \right)   + U(w,u,z) \,.
\end{equation}
The map~(\ref{CoordinatesWithOrientation}) still applies in presence of a scalar field, however now the two fixed-orientation shape spaces are two-dimensional planes, coordinatized by $(w_-,u_-) \in \mathbbm{R}^2$ and $(w_+,u_+) \in \mathbbm{R}^2$. This extends also to any number of additional fields: the shape space consists of two $N$-dimensional hyperplanes, one for each orientation.

We can now discuss one of the crucial steps allowing us to establish a continuation result: as we did in~\cite{ThroughTheBigBang,FlaviosInflationpaper,DavesScalarpaper}, we impose a particular topology on shape-space-with-orientation, which joins the borders of its two fixed-orientation connected components, making the overall space connected. This is done by compactifying shape space through the \emph{gnomonic projection:} each of the two fixed-orientation planes is mapped onto one of the hemispheres of a 2-sphere, with the origins mapped to the two poles, and the asymptotic borders mapped to the equator (see Fig.~\ref{GnomonicFig}). The gnomonic projection maps the coordinates $(w_\pm,u_\pm)$ into the spherical coordinates $\beta \in [0,\pi]$ and $\alpha \in [0,2\pi)$ as follows:
\begin{equation}
| \tan \beta | (\cos \alpha ,\sin \alpha) =
\left\{
\begin{aligned}
&(w_+,u_+)  \,, ~~ &\text{if } \beta < \pi/2 \,,
\\
&(w_-,u_-)  \,, ~~  &\text{if } \beta > \pi/2  \,.
\end{aligned}\right.
\end{equation}
In terms of these variables, and their conjugate momenta $p_\alpha$ and $p_\beta$, the Hamiltonian~(\ref{HamiltonianKS+Scalar2}) takes the following form:
\begin{equation}\label{HamiltonianOnShapeSphere}
H = \textstyle \frac 1 2  \cot^2 \beta   \, p_\alpha^2 +  \frac 1 2  \cos^4 \beta \, p_\beta^2    - \frac 1 2 
p_z^2   + U(\alpha,\beta,z) \,.
\end{equation}

\section{Continuation through the singularity}

In complete analogy with the cosmological models discussed in~\cite{ThroughTheBigBang,FlaviosInflationpaper,DavesScalarpaper},
 under the conditions described in Sec.~\ref{ScalarFieldSec} for the scalar field potential, the Hamiltonian on the shape sphere~(\ref{HamiltonianOnShapeSphere}) generates a dynamics that, near the singularity, asymptotes to that of a free point particle on the shape sphere (\emph{i.e.} a particle moving along great circles).
 The angle $\beta$ grows monotonically in this regime, and it can therefore be used as the independent variable, expressing the equations of motion in terms of derivatives of all the other variables with respect to $\beta$. In this formulation, all equations of motion except those for $p_\beta$ and $z$ are regular. However, the following change of variables:  
\begin{equation} \label{newdefs}
J = p_\beta \, \cos^2 \beta \,, \qquad v = z + \frac{ \tan \beta  \, p_z }{J} \,,
\end{equation}
gives a system of differential equations that are smooth at the singularity $\beta \to \left( \frac \pi 2 \right)^\pm$~\cite{FlaviosInflationpaper}:
\begin{equation}\label{FinalEqs}
\begin{aligned}
&\frac{d v}{d \beta} = - \frac{p_z \, p_\alpha^2}{\sin^2\beta J^3}
\,, \qquad
\frac{d \alpha}{d \beta} = \frac{p_\alpha}{\sin^2\beta J}
\,,
\\
&\frac{d J}{d \beta} = \frac{\cos \beta \, p_\alpha^2}{\sin^3 \beta J}
- \cos^2 \beta \frac{\partial U}{\partial \beta} \,,
\\
&\frac{d p_\alpha}{d \beta} = \frac{\partial U}{\partial \alpha} \,,
\qquad
\frac{d p_z}{d \beta} = \frac{\partial U}{\partial v} \,, 
\end{aligned}
\end{equation}
 where
\begin{equation}
 \begin{aligned}
 U (\alpha,\beta,v,J) &= - e^{\sqrt{\frac 8 3}  \left( v - \frac{ \tan \beta  \, p_z }{J}  \right) }  e^{- \sqrt{\frac 2 3} | \tan \beta | \cos \alpha } 
 \\
 &+ \frac{\kappa}{\nu^2} e^{\sqrt{6}  \left( v - \frac{ \tan \beta  \, p_z }{J}  \right)}   V\left({\textstyle \frac{|\tan \beta | \sin \alpha }{\sqrt{\kappa}}} \right) \,.
\end{aligned}   
 \end{equation}

Note that although $J$ is frequently in the denominator of these equations, $J \neq 0$ on solutions. Following from equation \ref{newdefs}, the only possibilities that allow for $J=0$ would be if the momentum $p_\beta$ were to vanish, or at the singularity where $\cos \beta$ can vanish. The former case is excluded dynamically as $\beta$ is increasing towards the singularity. At the singularity $p_\beta \rightarrow \infty$ such that $J$ remains finite and non-zero.
 
Just as in our previous results~\cite{ThroughTheBigBang,FlaviosInflationpaper,DavesScalarpaper}, Eqs.~(\ref{FinalEqs}) satisfy the assumptions of the existence and uniqueness theorem (the Picard-Lindel\"of theorem) for solutions of ordinary differential equations, and therefore, to each solution reaching the singularity from one hemisphere we can associate one and only one solution reaching the same point on the equator from the other hemisphere.

The Schwarzschild solution is a special case of the above system, in which there is no matter potential ($V=0$) and no scalar field momentum. In such a case it can be verified that $\alpha=p_\alpha=0$ is a solution to the equations of motion, which is represented by a great circle through the poles on the shape sphere. At the equator the solution continues along the great circle and crosses from one hemisphere to the other. On each hemisphere of shape space, the solution describes a black hole interior with either a left- or right-handed triad. The Picard-Lindel\"of theorem shows then that there is a unique continuation of the Schwarzschild interior beyond the singularity - it is an orientation-flipped interior of an otherwise identical black hole.

\section{Discussion}

\begin{figure}[t!]
\center
\includegraphics[width=0.5\textwidth]{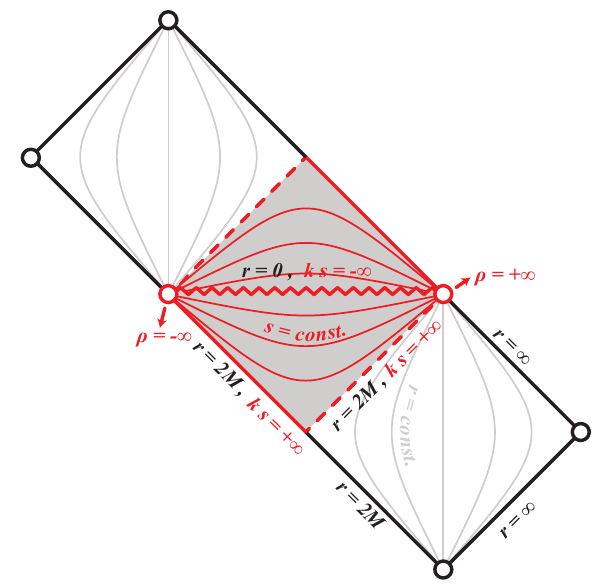}
\caption{The continuation of the Schwarzschild solution. At the singularity, the shape system remains well defined, and connects two Schwarzschild interiors described by right and left-handed triads.}\label{PenroseDiagramFig2}
\end{figure}

Our generalized dynamical system allows to continue singular solutions through the Schwarzschild singularity uniquely. As can be deduced by looking at the shape sphere in Fig.~\ref{GnomonicFig}, a great circle that crosses the equator won't be invariant under reflections with respect to the equator's plane (unless we're in the special case of a vertical, ``meridian'' circle). Then the solution continues to one that is objectively different: it is not simply the time-reversed repetition of the initial solution. After crossing the singularity, the shape degrees of freedom $w$ and $u$ will have a different evolution and will go through different pairs of values.

A legitimate question, at this point, is: what is the structure of the spacetime that corresponds to these continued solutions? The first thing we might investigate is its causal structure, which is entirely codified  in the evolution of the shape variable $w$. We know the causal structure associated to any half of each solution that is confined to one hemisphere: it is that of the region of Schwarzschild's spacetime that is inside the horizon: the shaded region in Fig.~\ref{PenroseDiagramFig1}. A full solution can then be associated to two such causal patches, and it is tempting to glue them at the singularity in the manner of Fig.~\ref{PenroseDiagramFig2}: one has two regions with opposite spatial orientations, looking like a black hole interior glued to a white hole interior. Extending these spacetimes beyond the horizons, one finds two asymptotically flat regions of opposite orientations, one in the causal past and one in the future.

This picture, however, is tentative and does not necessarily reflect actual physics. A Penrose diagram makes sense as an effective description of the causal relations between test particles propagating in a background spacetime, in a regime in which the backreaction of the particles on the geometry can be neglected. This is a reasonable assumption around most points in the Penrose diagram~\ref{PenroseDiagramFig2}, but not in the vicinity of the singularity. We cannot say, at the moment, what a test particle would experience upon crossing the singularity: that would need a dedicated analysis. Until that is done, we cannot be sure that timelike worldlines would behave smoothly at the singularity in the Penrose diagram~\ref{PenroseDiagramFig2}, and therefore the physical meaning of that diagram remains unclear.

This paper has shown how spacelike singularities at the center of black holes do not represent the end of the determinism of the solution. Together with \cite{ThroughTheBigBang,DavesScalarpaper,FlaviosInflationpaper}, this hints that the resolution of spacelike singularities may be a generic feature of the relational approach. However, this is far from the end of the problem of singularities. The Hawking-Penrose theorems still hold, and as yet it is not known how to extend geodesics beyond the singularity itself. Recent work \cite{Ashtekar2021}, see also \cite{Bianchi2018,DAmbrosio2018} has shown that despite these problems, given some extensions of spacetime beyond a singularity certain matter degrees of freedom can be deterministically evolved beyond these points. A tantalizing prospect is that relational descriptions may resolve the issues of singularities entirely classically. The ramifications for quantum gravity searches, many of which have their sights set on resolution of singularities, would be profound.

Another issue that should be investigated before proposing causal structures for our singularity-crossing solutions (and, in particular, before extending these structures outside of the horizons, is the fact that the Schwarzschild spacetime represents an eternal black hole, while realistic black holes are created through the collapse of matter.  This is better discussed within a matter collapse model that creates the black hole metric in its wake (\emph{e.g.} a thin-shell~\cite{ThinShell1,ThinShell2} or a Lemaitre--Tolman--Bondi model). Then, the study of the behaviour of the collapsing matter upon crossing the singularity should reveal the nature of the region beyond the singularity. A compelling possibility is that the singularity turns the collapse of the matter into an expansion, and the expanding matter leaves behind a pocket of spacetime with a white-hole metric.

\end{document}